\documentclass[conference]{IEEEtran}
\usepackage{tabularx}
\usepackage[pdftex]{graphicx} 
\usepackage{setspace}
\usepackage{xspace}
\usepackage{cite}
\usepackage{algorithm}
\usepackage{xcolor}
\usepackage{algorithmicx,algpseudocode}
\usepackage{amsmath}
\usepackage{comment}
\usepackage{url} 
\usepackage{makecell}
\usepackage{multirow} 

\usepackage{array, booktabs, caption}

\begin{document}

\title{Performance Evaluation of Scheduling Scheme \\ in O-RAN 5G Network using NS-3}



\author{\IEEEauthorblockN{
A. K. Subudhi\IEEEauthorrefmark{1},
A. Piccioni\IEEEauthorrefmark{2},
V. Gudepu\IEEEauthorrefmark{1},
A. Marotta\IEEEauthorrefmark{2}\IEEEauthorrefmark{3},
F. Graziosi\IEEEauthorrefmark{2},
R.M. Hegde\IEEEauthorrefmark{4}, 
K. Kondepu\IEEEauthorrefmark{1}
}
\IEEEauthorblockA{
\IEEEauthorrefmark{1} Indian Institute of Technology Dharwad, Dharwad, India;\\
\IEEEauthorrefmark{2} Università Degli Studi Dell'Aquila, L'Aquila, Italy; 
\IEEEauthorrefmark{3} WEST Aquila S.r.l., L'Aquila, Italy;\\
\IEEEauthorrefmark{4} Indian Institute of Technology Kanpur, Kanpur, India
\\
Email: ashit.subudhi.21@iitdh.ac.in}
}

\maketitle

\begin{abstract}

The integration of Open Radio Access Network (O-RAN) principles into 5G networks introduces a paradigm shift in how radio resources are managed and optimized.
O-RAN's open architecture enables the deployment of intelligent applications (xApps) that can dynamically adapt to varying network conditions and user demands. In this paper, we present radio resource scheduling schemes --- a possible O-RAN-compliant xApp can be designed. This xApp facilitates the implementation of customized scheduling strategies, tailored to meet the diverse Quality-of-Service (QoS) requirements of emerging 5G use cases, such as enhanced mobile broadband (eMBB), massive machine-type communications (mMTC), and ultra-reliable low-latency communications (URLLC).

We have tested the implemented scheduling schemes within an ns-3 simulation environment, integrated with the O-RAN framework. The evaluation includes the implementation of the Max-Throughput (MT) scheduling policy --- which prioritizes resource allocation based on optimal channel conditions, the Proportional-Fair (PF) scheduling policy --- which balances fairness with throughput, and compared with the default Round Robin (RR) scheduler. In addition, the implemented scheduling schemes support dynamic Time Division Duplex (TDD), allowing flexible configuration of Downlink (DL) and Uplink (UL) switching for bidirectional transmissions, ensuring efficient resource utilization across various scenarios. The results demonstrate resource allocation's effectiveness under MT and PF scheduling policies. To assess the efficiency of this resource allocation, we analyzed the Modulation Coding Scheme (MCS), the number of symbols, and Transmission Time Intervals (TTIs) allocated per user, and compared them with the throughput achieved. The analysis revealed a consistent relationship between these factors and the observed throughput.

$\textit{Index Term}$-- 5G Network, ns3-O-RAN, PF, MT, RR, resource allocation, scheduling, MCS


\end{abstract}

\section{Introduction}

Fifth Generation (5G) networks represent a significant leap forward from the Long Term Evolution (LTE) standard developed by 3rd Generation Partnership Project (3GPP), bringing enhanced connectivity, ultra-low latency, and higher data rates to support the ever-growing demand for mobile broadband services~\cite{rec2020itu}.
5G networks are designed to accommodate a variety of use cases --- enhanced mobile broadband (eMBB), massive machine-type communications (mMTC), and ultra-reliable low-latency communications (URLLC) --- enabling seamless experiences for bandwidth-intensive applications like augmented reality (AR), virtual reality (VR), real-time gaming, and high-definition video streaming.

One of the key features of 5G is it's flexible bandwidth usage, supporting a wide range of frequency bands from sub-6 $GHz$ to millimeter waves (mmWave), allowing network operators to optimize their spectrum usage and deliver improved services globally.
5G networks also aim to enhance Quality of Service (QoS) support through advanced Radio Resource Management (RRM) techniques that surpass those used in 4G~\cite{monikandan2017review}.
In the 5G system architecture, where next-generation NodeBs (gNBs) handle packet scheduling and RRM, efficient resource scheduling is vital for optimal system performance.
However, existing algorithms like Round Robin have certain limitations~\cite{centofanti2024end} --- offers fairness by equally allocating resources without considering channel conditions, leading to inconsistent throughput.  
A work in~\cite{ala2022scheduling}, presents adaptive scheduling for Centralized-RAN (C-RAN) based on the user traffic demands over a particular region and allocation of Radio Units (RUs) to meet the service demands.
The traditional static scheduling methods have several challenges to adapt with the dynamic service demands of the 5G systems.

\begin{figure*}[t!]
    \centering
    \includegraphics[width=0.7\textwidth]{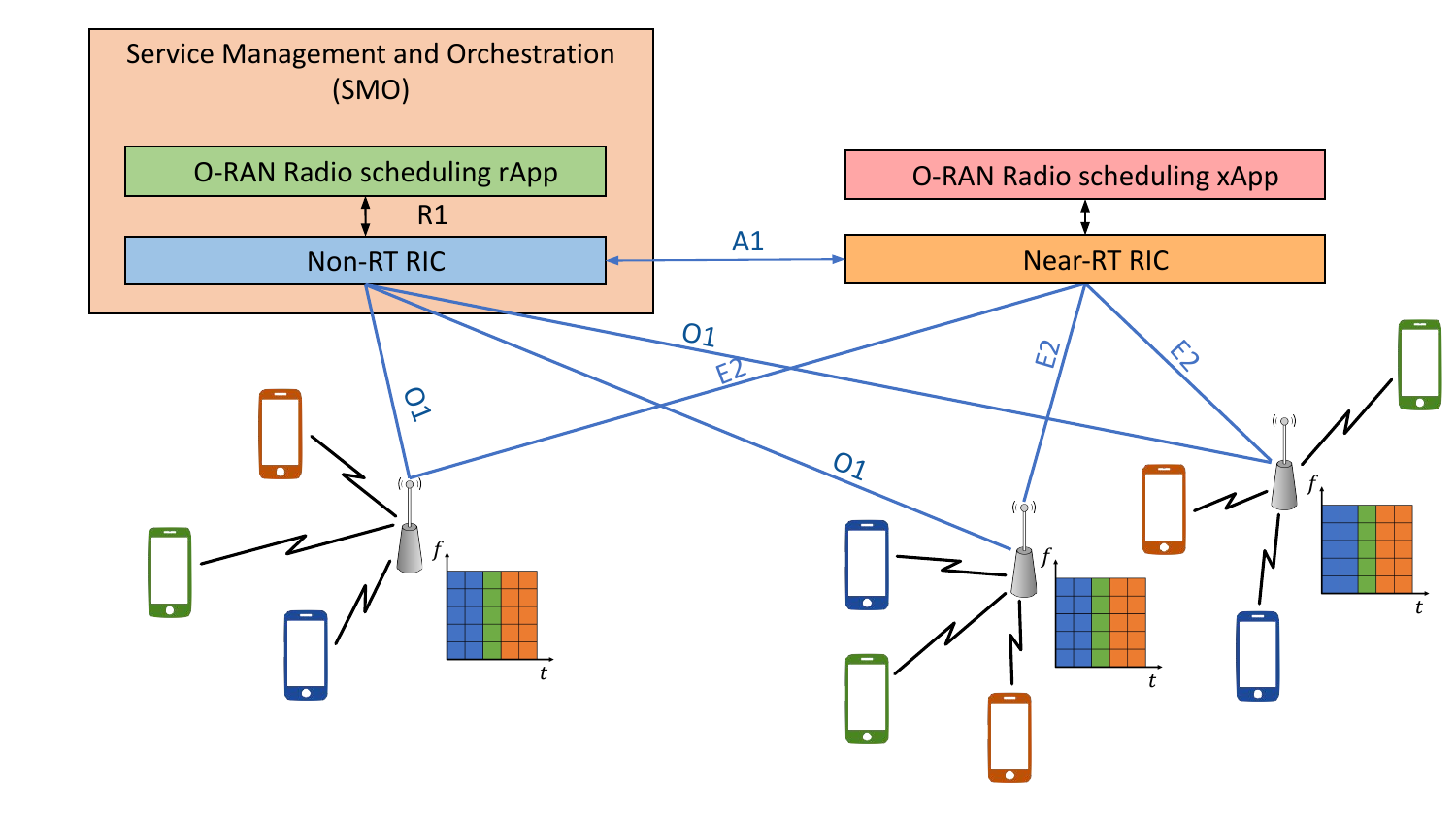}
    \caption{System Model}
    \vspace{-0.5cm}
    \label{fig:sysmod}
\end{figure*}

However, to configure the scheduling policies on-the-fly based on the network conditions, the advent of Open Radio Access Network (Open RAN) technology is bringing more flexibility, interoperability, and intelligent solutions into the 5G and Beyond (B5G) networks.
In addition, O-RAN enhances traditional scheduling frameworks by integrating AI/ML capabilities and insights into RAN user data traffic through the RAN Intelligent Controller (RIC), which operates in both Near- and Non-Real Time (Near- and Non-RT). 
These controllers manage the RAN through Non-RT Application (rApp) and Near-RT Application (xApp), offering more granular control and optimization of resource allocation at different operational levels~\cite{gudepu2024drift}.

In the O-RAN environment, the efficiency of the scheduler is augmented by the RIC, which enables the dynamic adaptation of scheduling policies based on real-time data and predictions. This approach allows for more responsive and adaptive scheduling that can cater to the diverse needs of 5G applications, including eMBB, mMTC, and URLLC, while addressing the limitations of existing algorithms. The open and disaggregated nature of O-RAN fosters innovation in scheduling algorithms, driving the evolution of more sophisticated and intelligent RRM techniques. These innovations are crucial for optimizing system performance metrics such as \emph{error rate}, \emph{latency}, \emph{fairness}, \emph{user throughput}, and \emph{overall system throughput}, particularly as 5G networks continue to evolve to meet the demands of future applications and services~\cite{centofanti2023end}.
~\cite{arslan2023dynamic}, explores the Federated learning-based Deep Reinforcement Learning (FDRL) for the Medium Access Control (MAC) layer for adaptive scheduling in O-RAN. 
However,~\cite{arslan2023dynamic} is not focusing on any x/rApp implementation which is essential to configure the RAN scheduling policies dynamically based on the network conditions through E2 control messages from Near-RT RIC and A1 policies from Non-RT RIC.

The focus of the proposed work is to implement an xApp for radio scheduling, which enables the deployment of flexible strategies, and testing it within an ns-3 environment integrated with O-RAN, i.e. ns-O-RAN~\cite{nsoran}. 
The proposed approach leverages the flexibility and intelligence of O-RAN's RICs to dynamically optimize resource allocation and enhance overall network performance.

The rest of the paper is organized as follows. Section II describes the O-RAN framework. Section III describes the scenario established using ns3 with O-RAN (ns-O-RAN) and the MAC scheduling policies. Section IV explains the experimental results. Finally, Section VI concludes the paper with essential observations.

\section{System Model}
We consider the system model depicted in Fig.~\ref{fig:sysmod}, where a generic RAN architecture is designed to provide connectivity through efficient radio resource scheduling. The overall architecture leverages the O-RAN framework, which introduces openness, flexibility, and interoperability to traditional RANs. The system consists of several components, among which the Non-RT RIC, the Near-RT RIC, which are interconnected through the A1 interface, as well as the Service and Management Orchestration (SMO) framework. The SMO plays a crucial role in managing the life-cycle of network functions and applications within the O-RAN ecosystem.

The proposed architecture supports a set of heterogeneous services, ranging from high-throughput data transmission to low-latency communications. This varying demand is managed by exploiting an ad-hoc xApp that is responsible for the efficient and dynamic scheduling of radio resources among the UEs. 
The xApp continuously monitors network status, channel conditions, and UE requirements to dynamically adjust the resource allocation and optimize the network performance. 
Furthermore, the xApp can operate in geographical areas involved in specific situations or events that would benefit from dynamic scheduling, for instance implementing more multimedia-oriented scheduling strategies (e.g., multi-cast or broadcast) near a stadium. In this model, we assume that the UEs are connected to multiple gNBs. 
The xApp can enforce different radio scheduling strategies, e.g., Round Robin~\cite{4626953}, Max Throughput (MT)~\cite{5757800}, or Proportional Fair~\cite{Femenias2017}, as well as different multiple access techniques, e.g., Time-Division Multiple Access (TDMA), or Orthogonal Frequency-Division Multiple Access (OFDMA). 

The xApp operates within a broader ecosystem, collaborating with a purposely designed rApp for radio scheduling and other network functions through the SMO. The rApp provides a higher-level control by defining the optimization strategies that the xApp must follow. 
This approach allows the architecture to maintain high efficiency in real-time operations. 
The xApp relies on information from the Near-RT RIC, which is responsible for the optimal selection of the radio scheduling strategy. 
It receives critical inputs about the network condition, traffic pattern, and specific parameters required by the scheduling techniques, including bandwidth parameters per service (i.e., adopted numerology, sub-carrier spacing, etc.,), Channel Quality Indicator (CQI), or SNR thresholds depending on the target BLock Error Rate (BLER), just to name a few. In addition, the rApp can collect information from external systems, which translates into policies provided to the xApp. This allows to dynamically variate scheduling strategies over time based on network conditions as well as external conditions such as vehicular traffic forecasts, high-density events, and critical disaster scenarios.

\section{Experimental Setup}

\begin{figure}
    \centering
    \includegraphics[width=1.0\columnwidth]{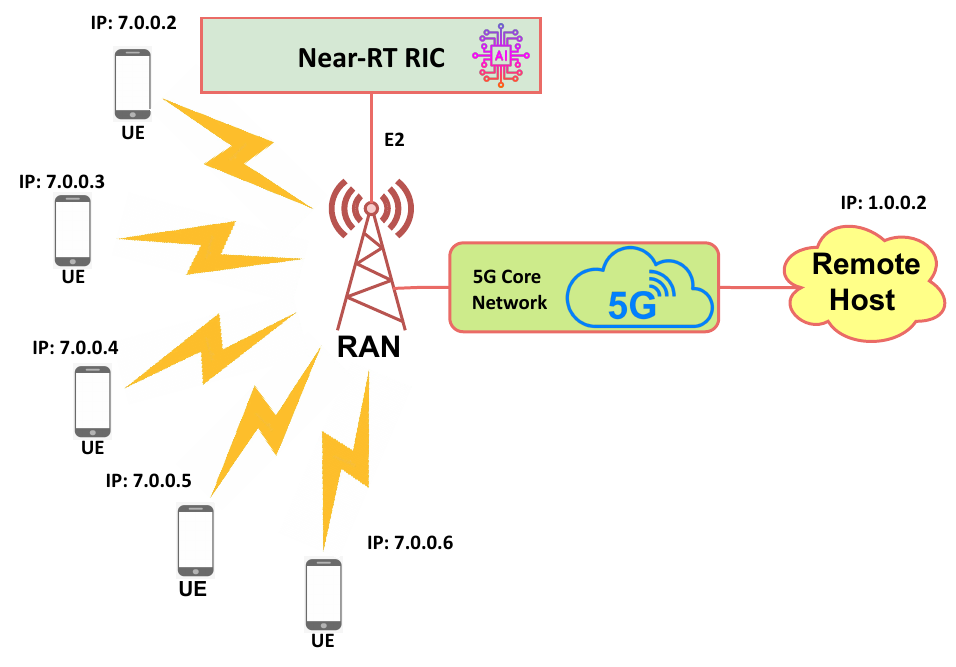}
    \caption{A Scenario in 5G Network using ns-3}
    \label{fig:my_label3b}
\end{figure}

In ns3, we have set up a 5G scenario where UEs are connected wirelessly to a gNB. The gNB, in turn, is connected to the 5G Core Network (5G CN), a crucial component of the CN architecture that connects to the remote host or internet. 
In addition, the gNB is connected to the Near-RT RIC via the E2 interface. A communication channel is established between the gNB, Packet Data Network (PDN) Gateway (PGW), and the remote host. The Network Interface Cards (NICs) with software drivers, known as network devices, are installed in the PGW, UEs, and gNB to enable network connectivity. 
The topology helper class facilitates connecting these network devices to nodes and channels while assigning IP addresses. 
The UEs in this scenario are considered to be non-mobile. Fig. \ref{fig:my_label3b} illustrates our 5G O-RAN scenario built using ns-3.

The protocol stack implemented in both the UEs and gNB includes multiple layers: the physical layer, Medium Access Control (MAC) layer, Radio Link Control (RLC) layer, Packet Data Convergence Protocol (PDCP) layer, and Radio Resource Control (RRC) layer. The Data transmission primarily occurs in Downlink (DL) and Uplink (UL). In the DL, data flows from the remote host to the PGW, then to the gNB, and finally reaches the UEs. 

The data traverses several protocol layers during transmission from the PGW to the gNB: (i) The packets first pass through the PDCP layer, where header compression and encryption are applied; (ii) The packet goes through the RLC layer --- where data segmentation and in-order delivery are ensured --- and then forwarded to the MAC layer; (iii) The packets are scheduled in the MAC layer to manage radio resources; and (iv) Finally, the physical layer transmits the data over the air, carrying all information from the MAC transport channels. 

Upon reception by the UE physical layer, the data is forwarded to the upper layers and delivered to the appropriate application or process on the UE. Similarly, in the UL, data is transmitted from the UE to the gNB, then to the PGW, and finally reaches the remote host.

\begin{figure}
    \centering
    \includegraphics[width=1.0\columnwidth]{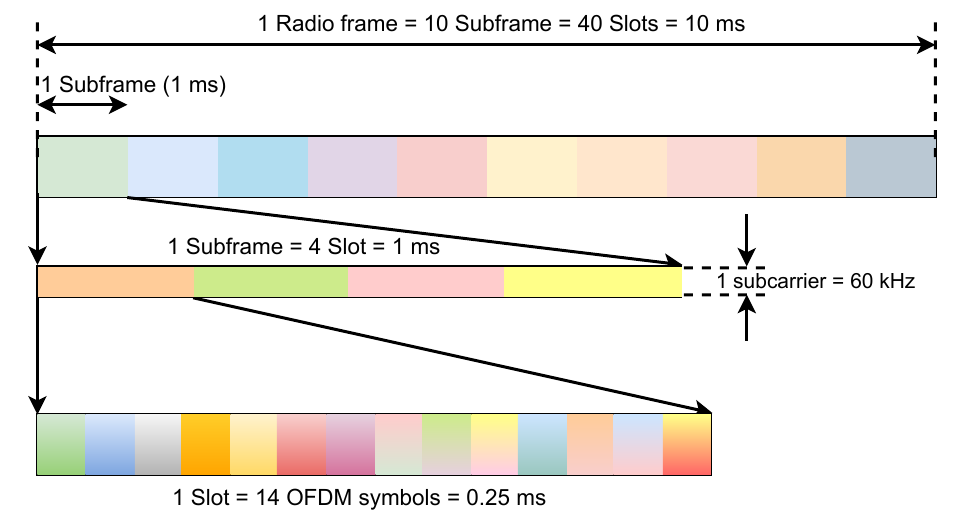}
    \caption{5G NR Frame Structure}
    \label{fig:my_label3c}
\end{figure}

The UL/DL data flow is highlighted through the 5G NR frame structure~\cite{etsi2019} as shown in Fig.~\ref{fig:my_label3c}. The frame structure is one of the key elements to perform various scheduling policies applied in the MAC layer of the gNB in order to allocate the radio resources efficiently. 
As per the 3GPP standard, different types of numerology ($\mu$) are proposed, which leads to different frame structures. Fig.~\ref{fig:my_label3c} explains the 5G NR frame structure for $\mu$=2 with the subcarrier spacing of 60 KHz. A frame of $10ms$ consists of 10 subframes, and each subframe of $1ms$ consists of a certain number of slots, which is calculated based on the numerology value, i.e. $2^{\mu}$, and each slot consists of 14 OFDM symbols. These OFDM symbols carry the \emph{data}, \emph{control information}, or \emph{reference signals}. Under the flexible slot structure in NR, the first OFDM symbol in a slot is designated for DL control, and the last OFDM symbol is reserved for UL control. The remaining OFDM symbols within the slot can be dynamically allocated to either DL or UL data, facilitating dynamic TDD. This structure allows for a flexible and configurable slot design, enabling rapid DL-UL switching for bidirectional transmissions~\cite{patriciello2019e2e}.  

To accommodate varying user demands, we implement different scheduling policies. The different scheduling policies are followed in the MAC layer such as RR scheduling, MT scheduling, and PF scheduling policy to allocate the resources. RR scheduling is a time-insensitive policy that distributes radio resources to users without considering current channel conditions~\cite{4626953}. 
However, RR enhances fairness among users in terms of resource allocation, thus, negatively impact overall system performance. 
The MT scheduling policy allocates radio resources based on channel conditions. The gNB collects CQI reports from users and prioritizes those with the highest CQI values, indicating the best channel conditions. However, this technique tends to disadvantage users far away from the gNB (i.e. at the cell edge), as they typically experience poorer performance. Consequently, the scheduling policy often favours users with the highest CQI, which can lead to reduced resource allocation for other users~\cite{5757800}. 

\begin{figure}
    \centering
    \includegraphics[width=0.50\textwidth]{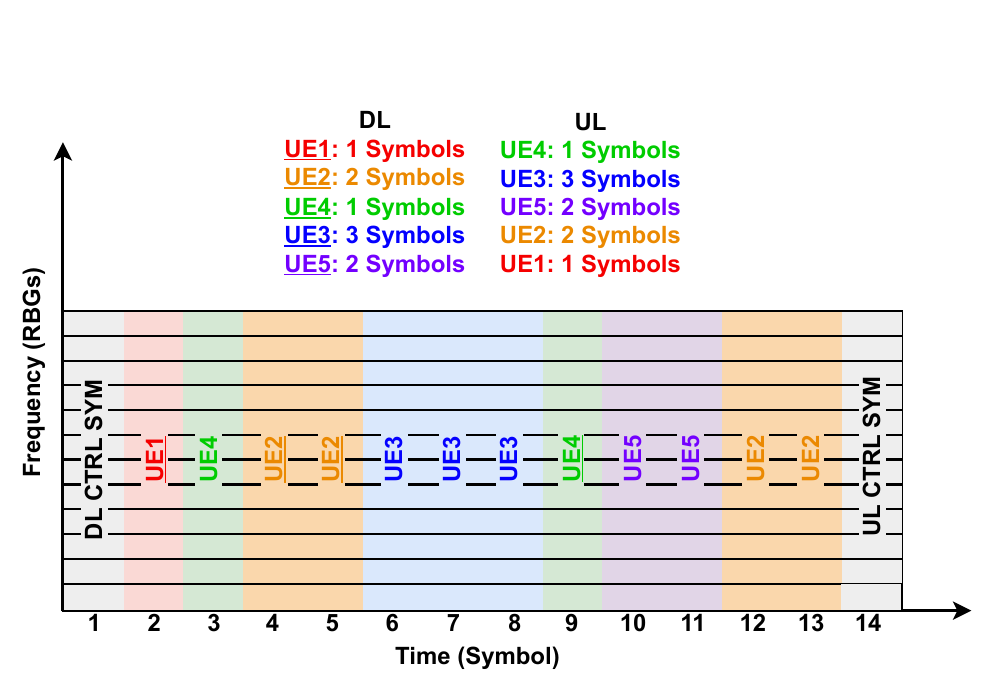}
    \caption{Example of User Allocation using MT Scheduling}
    \label{fig:my_label3d}
\end{figure}

Fig.~\ref{fig:my_label3d} illustrates the MT scheduling policy for 5 UEs in both DL and UL data transmission within a specific slot of a subframe. 
The UE demands are illustrated in terms of symbols, with resources allocated based on the CQI values in both DL and UL, considering the entire frequency range of Resource Block Groups (RBGs) (according to channel bandwidth) for each slot.  

In contrast, the PF scheduling policy strikes a balance between fairness and channel conditions. It allocates resources by considering both the users channel quality and fairness, ensuring that even users with weaker channel conditions are fairly treated. This policy leads to more equitable resource distribution while still maintaining high throughput, providing a compromise between the extremes of RR and MT scheduling. The effectiveness of this resource allocation will be assessed using performance evaluation parameters.

To assess the performance of the MAC scheduler's radio resource allocation, we need to analyze various network performance parameters, including throughput and delay. The definitions of the considered performance metrics are listed as follows: 

\begin{itemize}
    \item Throughput: Throughput measures the effective data transmission capacity of a network. It is calculated as the number of bits received in packets divided by the total delivery time. It reflects how efficiently data is transmitted over the network~\cite{9773317}.\\
    \begin{equation*} 
     Throughput=\frac {\sum {Rx\_Packet\_Size}}{Delivery\_Time} \tag{1}
     \end{equation*}
     where, $Rx\_Packet\_Size$ denotes the size of the received packets, and $Delivery\_Time$ represents the time to deliver the data to its destination.

    \item Delay: Delay measures the time a packet takes to reach the end-user. This performance parameter quantifies the perceived latency in successfully receiving the packets~\cite{9773317}.
   \begin{equation*} 
    Delay=T_{r}-T_{t} \tag{2}
    \end{equation*}
    where, $T_{r}$ represents the time the data recieved and the $T_{t}$ represents the time the data transmitted.
 
\end{itemize}

\section{Experimental Results}

This section outlines the evaluation of the performance parameters of the scenario built in the ns-O-RAN environment. The performance and effectiveness of the resource allocation policies have been extensively analyzed and evaluated. We examine how each scheduling policy affects the network performance, considering a variety of user demands. Table~\ref{tab:my_label} presents the details on the different simulation parameters of the described scenario.
 
\begin{table}[]
    \centering
\begin{tabularx}{\columnwidth}{| X | X |} 
\hline
\textbf{Parameter} & \textbf{Value}  \\
\hline
Channel bandwidth & 20 MHz  \\ 
\hline
Frame duration & 10 ms \\
\hline
Numerology $(\mu)$ & 2 \\
\hline
Frequency & 3.5 GHz \\
\hline
No. of slots per subframe & 4 \\
\hline
No. of OFDM symbols/slot & 14 \\
\hline
Subcarrier frequency & 60 KHz \\
\hline
Packet size & 1000 Bytes \\ 
\hline
Number of gNB & 1 \\
\hline
Number of UEs & 1-10 \\
\hline
Experimental duration & 12000 TTIs\\
\hline
Types of user demand & 3\\
\hline
Channel model & 3GPP Umi Street Canyon\\
\hline
Modulation Schemes & 64 QAM\\
\hline
\end{tabularx}
    \caption{Simulation Parameters}
    \label{tab:my_label}
\end{table}



\begin{figure}[H]
    \centering
    \includegraphics[width=1.1\columnwidth]{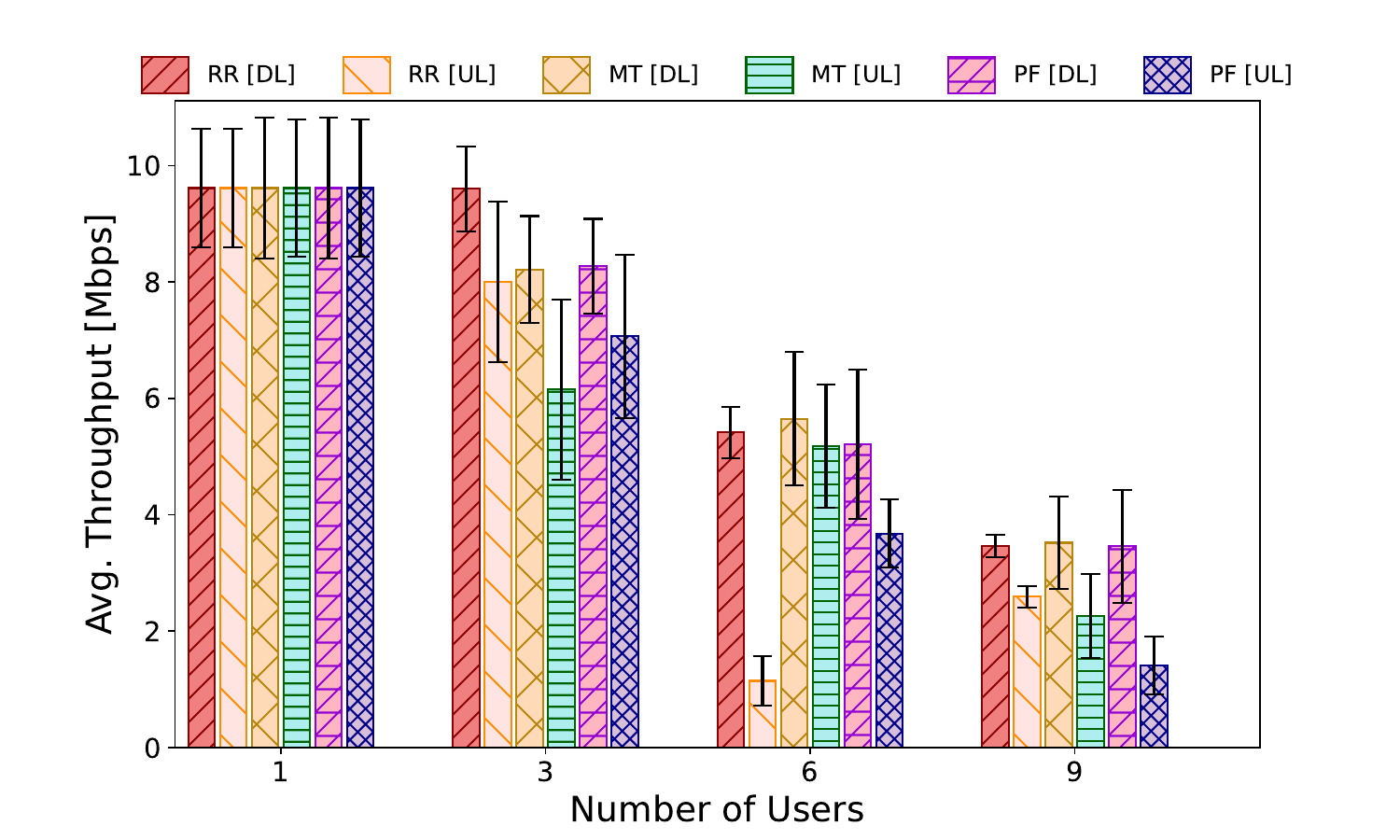}
    \caption{Average Throughput vs Number of Users}
    \label{fig:my_label3e}
\end{figure}

Fig.~\ref{fig:my_label3e} illustrates the average throughput as a function of the number of users. This average throughput is measured for the RR, MT and PF scheduling policy with equal weights given to both DL and UL users. 
As expected, throughput decreases with the number of users increase. This decrease is attributed to the scheduling policy's inability to adequately meet the demands of each user within a single slot. Consequently, these scheduling policies exhibit similar trends aligning with the standards~\cite{1543653}. However, while the average throughput remains comparable across the different schemes as the number of users increases, the level of fairness among users differs significantly, which is a critical factor that directly impacts the use cases in the 5G network.

\begin{figure}
    \centering
    \includegraphics[width=1.1\columnwidth]{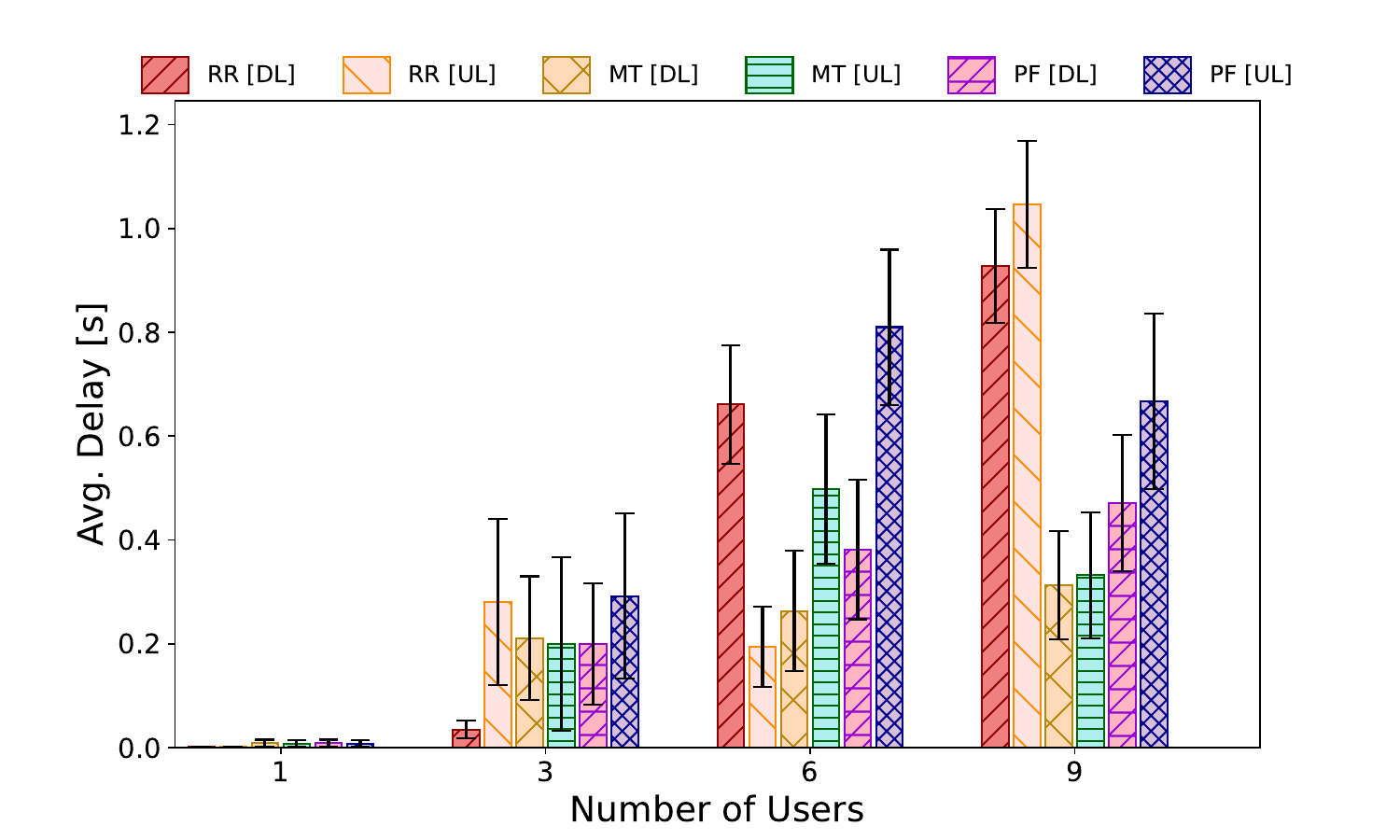}
    \caption{Average Delay vs Number of Users}
    \label{fig:my_label3f}
\end{figure}

Fig.~\ref{fig:my_label3f} illustrates the average delay in relation to the number of users for RR, MT and PF scheduling policies. The figure demonstrates that as the number of users increases, the average delay rises in all the policies. In the RR policy, which emphasizes fairness in resource allocation, the addition of more users in both DL and UL results in partial resource allocation for each user, leading to longer wait times. As user numbers grow, competition for resources intensifies, further increasing the overall delay. In contrast, the delay in MT scheduling is generally lower than in RR, as MT prioritizes users with the best channel conditions, allowing for more efficient resource utilization and quicker data transmission for those users. Meanwhile, the PF scheduling policy strikes a balance, resulting in a moderate delay compared to the other policies, as it considers both fairness and throughput maximization.

\begin{figure}
    \centering
    \includegraphics[width=1.05\columnwidth]{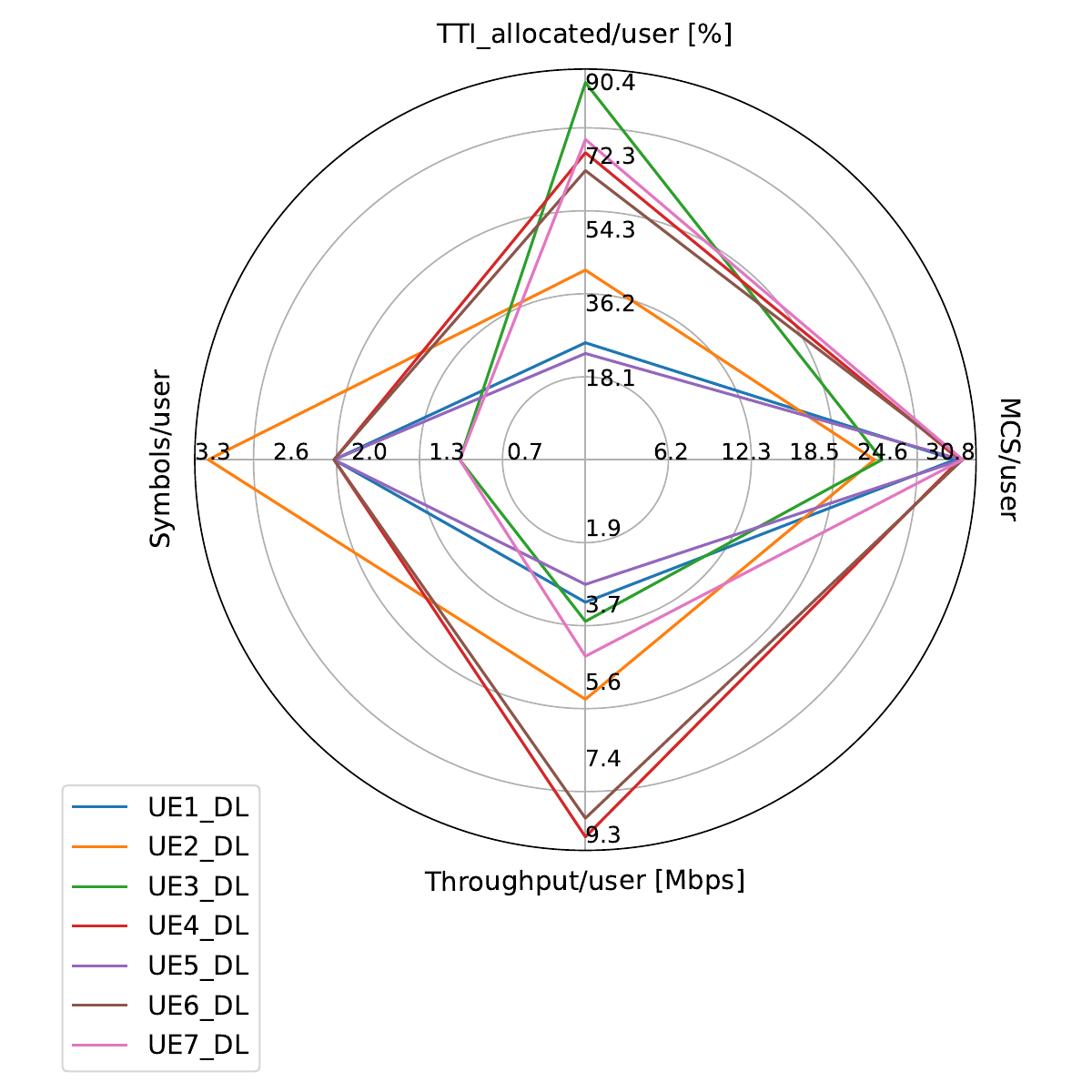}
    \caption{Variation in DL Throughput per User in PF Scheduling Policy }
    \label{fig:my_label3i}
\end{figure}

\begin{figure}
    \centering
    \includegraphics[width=1.05\columnwidth]{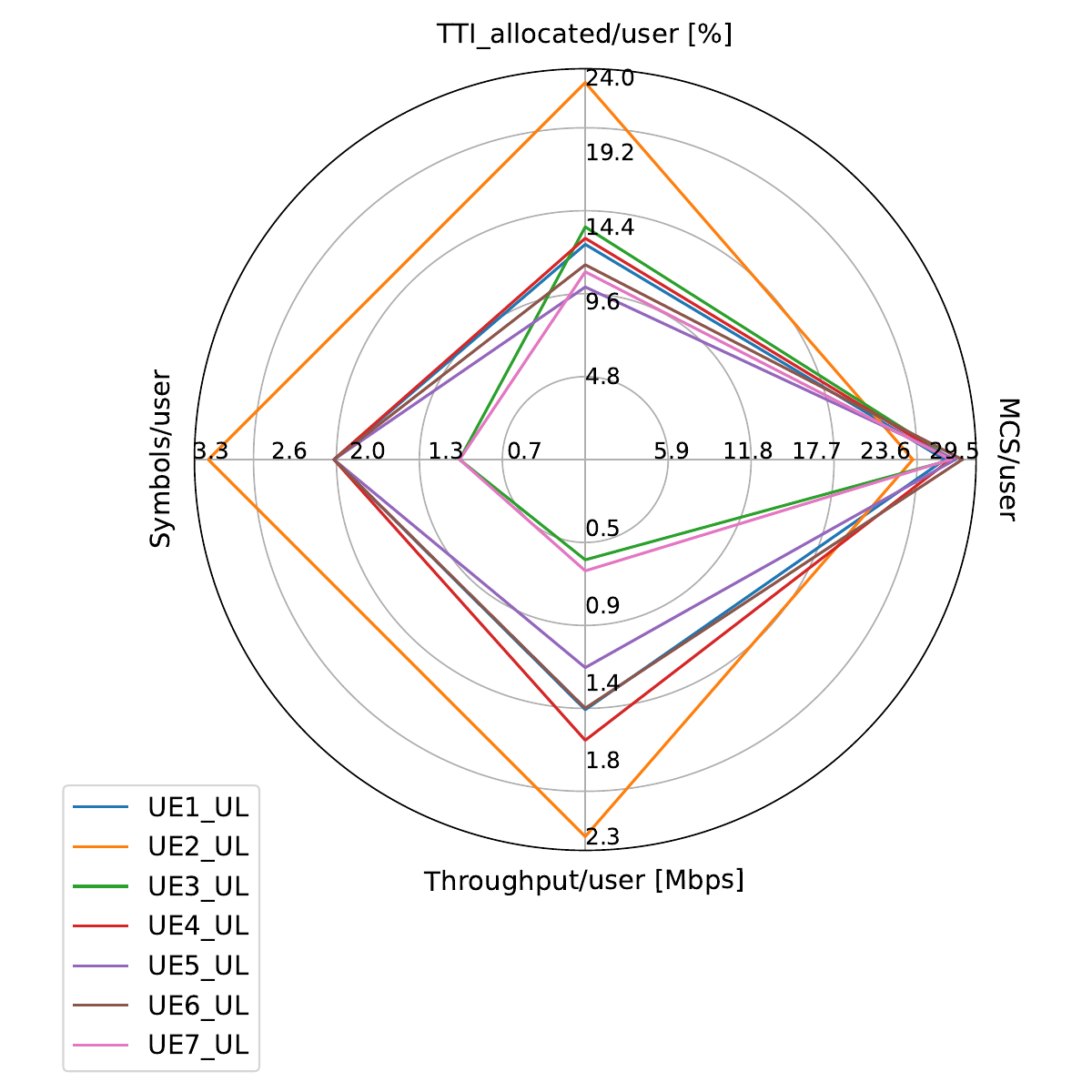}
    \caption{Variation in UL Throughput per User in PF Scheduling Policy}
    \label{fig:my_label3j}
\end{figure}


Fig.~\ref{fig:my_label3i} and Fig.~\ref{fig:my_label3j} illustrate the variation in throughput for both DL and UL users in the PF scheduling policy. Here, the user throughput is primarily influenced by three different parameters: (i) Modulation and Coding Scheme (MCS) --- the number of useful bits that can be carried by a single symbol.
(ii) Transmission Time Interval (TTI) allocation --- the smallest unit of time in the radio frame, used by gNB for scheduling the users for both DL and UL transmission. 
(iii) Number of symbols required --- the requirement of number symbols is fixed by each user as per the service they use. During the simulation, 3 different user demands are considered, and each user is pre-associated with particular demands ranging from 1 to 3 symbols per user.


Fig.~\ref{fig:my_label3i} shows the DL throughput in the PF scheduling policy for the considered 7 users. The UE4 and UE6 experience higher throughput, around 8 Mbps. This increased throughput can be attributed to the higher received MCS (i.e., 28 MCS), higher TTI allocation (\%) and the moderate requested symbol/user. In contrast, for UE1, UE2, UE3, UE5 and UE7, the throughput is around 3 Mbps and 6 Mbps, respectively. Although their MCS values are also high (28 MCS), the TTI allocation (\%) is medium to high and the requested symbol/user is of low to high, which impacts their overall throughput. Similarly, in Fig.~\ref{fig:my_label3j} we observe the UL throughput in the PF scheduling policy for the 7 users based on variation in each performance metric.

These observations clearly demonstrate the effectiveness of the radio resource allocation in the PF scheduling policy. They highlight how the policy successfully balances both fairness and throughput, ensuring equitable resource distribution while optimizing user performance.



\section{Conclusion}

In this paper, we evaluated scheduling policies for resource allocation based on user demands in 5G networks within an O-RAN framework, utilizing the ns-3 simulation environment. We compared the performance of the PF and MT scheduling policies with the RR scheduling policy. The PF scheduling policy prioritizes users based on channel conditions and fairness, resulting in higher throughput. The MT scheduling policy also prioritizes users according to channel conditions, leading to enhanced throughput. In contrast, the RR policy ensures fairness by evenly distributing resources among all users, regardless of their channel conditions. While this approach promotes equitable access, it may lead to less optimal overall network performance, especially in scenarios with varying channel conditions. Incorporating O-RAN principles into 5G networks enhances flexibility and adaptability in scheduling policies guided by the RIC. This integration enables dynamic and intelligent scheduling strategies that better address the demands of diverse use cases and network conditions.

\section*{Acknowledgments}
This work has been partially supported by TTDF “SMART-RIC6G: Smart Drift-Handling
Enabler for RAN Intelligent Controllers in 6G Networks (TTDF/6G/422)” project. This work has been partially supported by the European Union through SNS Joint Undertaken SEASON project (GA: 101096120), European Union - NextGenerationEU under the Italian Ministry of University and Research (MUR) National Innovation Ecosystem grant ECS00000041 - VITALITY - CUP E13C22001060006, and by the European Union under the Italian National Recovery and Resilience Plan (NRRP) of NextGenerationEU, partnership on “Telecommunications of the Future” (PE00000001 - program “RESTART”)

\bibliographystyle{IEEEtran}
\bibliography{FNWF_2024_Sub}

\end{document}